# Survey on Improved Scheduling in Hadoop MapReduce in Cloud Environments

B.Thirumala Rao
Associate Professor
Dept. of CSE
Lakireddy Bali Reddy College of Engineering

Dr. L.S.S.Reddy
Professor & Director
Dept. of CSE
Lakireddy Bali Reddy College of Engineering

## ABSTRACT

Cloud Computing is emerging as a new computational paradigm shift. Hadoop-MapReduce has become a powerful Computation Model for processing large data on distributed commodity hardware clusters such as Clouds. In all Hadoop implementations, the default FIFO scheduler is available where jobs are scheduled in FIFO order with support for other priority based schedulers also. In this paper we study various scheduler improvements possible with Hadoop and also provided some guidelines on how to improve the scheduling in Hadoop in Cloud Environments.

## 1. INTRODUCTION

Cloud computing [1] refers to the use of shared computing resources to deliver computing as a utility, and serves as an alternative to having local servers handle computation. Cloud computing groups together large numbers of commodity hardware servers and other resources to offer their combined capacity on an on-demand, pay-as-you-go basis. The users of a cloud have no idea where the servers are physically located and can start working with their applications. This is the primary advantage of cloud computing which distinguishes it from grid or utility computing. The primary concept behind Cloud Computing isn't a new idea. John McCarthy within the sixties imagined that "processing amenities is going to be supplied to everyone just like a utility". The word "cloud" has already been utilized in numerous contexts such as explaining big ATM systems within the 1990s. Nevertheless, it had been following Google's BOSS Eric Schmidt utilized the term to explain the company type of supplying providers over the Web within 2006. Since then, the term "cloud computing" has been used mainly as a marketing term. Lack of a standard definition of cloud computing has generated a fair amount of uncertainty and confusion. For this reason, significant work has been done on standardizing the definition of cloud computing. There are over 20 different definitions from a variety of sources. In this paper, we adopt the definition of cloud computing provided by The National Institute of Standards and Technology (NIST), as it covers, in our Opinion, all the essential aspects of cloud computing:

**NIST definition of cloud computing[2]: "***Cloud computing is a model for enabling convenient, on-demand network access to a shared pool of configurable computing resources (e.g.,* networks, servers, storage, applications, and services) that can be rapidly provisioned and released with minimal management effort or service provider interaction".

Cloud computing concept is motivated by latest data demands as the data stored on web is increasing drastically in recent times. The computing resources (e.g. servers, storage and services) in a cloud can automatically be scaled up to meet the dynamic demands of users by its virtualization and distributed system technology. In addition to that, it also provides redundancy and backup features to overcome the hardware failure problems. In cloud environments data processing has become an important research problem. As cloud is a proper distributed system platform, parallel programming model like *MapReduce [4]* is widely used for developing scalable and fault tolerant applications deployable on cloud. Rest of the paper is organized as follows: In section 2 Hadoop is summarized and various current schedulers are discussed in section 3. Hadoop scheduler improvements are discussed in section 4. Finally we conclude with discussion of future work in section 5.

## 2. HADOOP

Hadoop has been successfully used by many companies including AOL, Amazon, Facebook, Yahoo and New York Times for running their applications on clusters. For example, AOL used it for running an application that analyzes the behavioral pattern of their users so as to offer targeted services. Apache Hadoop [3] is an open source implementation of the Google's MapReduce [4] parallel processing framework. Hadoop hides the details of parallel processing, including data distribution to processing nodes, restarting failed subtasks, and consolidation of results after computation. This framework allows developers to write parallel processing programs that focus on their computation problem, rather than parallelization issues. Hadoop includes 1) Hadoop Distributed File System (HDFS): a distributed file system that store large amount of data with high throughput access to data on clusters and 2) Hadoop MapReduce: a software framework for distributed processing of data on clusters.

## 2.1 HDFS- Distributed file system

Google File System (GFS) [5] is a proprietary distributed file system developed by Google and specially designed to provide efficient, reliable access to data using large clusters of commodity servers. Files are divided into chunks of 64 MB, and are usually appended to or read and only extremely rarely overwritten or shrunk. Compared with traditional file systems, GFS is designed and optimized to run on data centers to provide extremely high data throughputs, low latency and survive individual server failures. Inspired by GFS, the open source Hadoop Distributed File System (HDFS) [6] stores large files across multiple machines. It achieves reliability by replicating the data across multiple servers. Similarly to GFS, multiple replicas of data are stored on multiple compute nodes to provide reliable and rapid computations. Data is also provided over

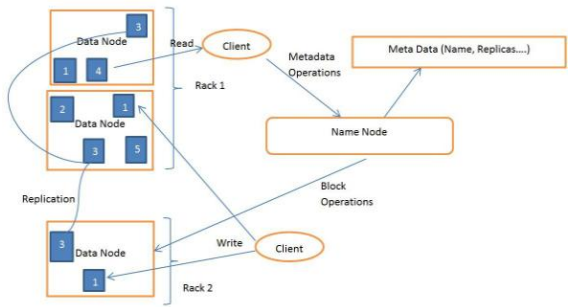

Fig1: Hadoop Distributed File System [6]

HTTP, allowing access to all content from a web browser or other types of clients. HDFS has master/slave architecture.

As shown in Fig.1 HDFS consists of a single NameNode and multiple DataNodes in a cluster. NameNode is responsible for mapping of data blocks to DataNodes and for managing file system operations like opening, closing and renaming files and directories. Upon the instructions of NameNode, DataNodes perform block creation, deletion and replication of data blocks. The NameNode also maintains the file system namespace which records the creation, deletion and modification of files by the users. NameNode decides about replication of data blocks. In a typical HDFS, block size is 64MB and replication factor is 3 (second copy on the local rack and third on the remote rack).

## 2.2 Hadoop MapReduce Overview

MapReduce is one of the parallel data processing paradigm designed for large scale data processing on cluster-based computing architectures. It was originally proposed by Google to handle large-scale web search applications. This approach has been proved to be an effective programming approach for developing machine learning, data mining, and search applications in data centers. Its advantage is that it allows programmers to abstract from the issues of scheduling, parallelization, partitioning, replication and focus on developing their applications. As shown in Fig.2 Hadoop MapReduce programming model consists of data processing functions: *Map* and *Reduce*. Parallel Map tasks are run on input data which is partitioned into fixed sized blocks and produce intermediate output as a collection of <key, value> pairs. These pairs are shuffled across different reduce tasks based on <key, value>

pairs. Each Reduce task accepts only one key at a time and process data for that key and outputs the results as <key, value> pairs. The Hadoop MapReduce architecture consists of one JobTracker (Master) and many TaskTrackers (Workers). The JobTracker receives job submitted from user, breaks it down into map and reduce tasks, assigns the tasks to Task Trackers, monitors the progress of the Task Trackers, and finally when all the tasks are complete, reports the user about the job completion. Each Task Tracker has a fixed number of map and reduce task slots that determine how many map and reduce tasks it can run at a time. HDFS supports reliability and fault tolerance of MapReduce computation by storing and replicating the inputs and outputs of a Hadoop job. Since Hadoop jobs have to share the cluster resources, a scheduling policy is used to determine when a job can execute its tasks. Earlier Hadoop had a very simple scheduling algorithm operates on First-in First-out (FIFO) basis for scheduling users' jobs by default. Later significant amount of research took place in developing more effective and environment-specific schedulers. All those schedulers were discussed in the next section.

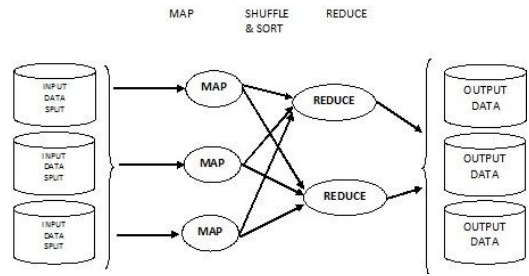

Fig 2: Hadoop MapReduce

## 3. SCHEDULING IN HADOOP

The default Scheduling algorithm is based on FIFO where jobs were executed in the order of their submission. Later on the ability to set the priority of a Job was added. Facebook and Yahoo contributed significant work in developing schedulers i.e. Fair Scheduler [7] and Capacity Scheduler [8] respectively which subsequently released to Hadoop Community.

## 3.1 Default FIFO Scheduler

The default Hadoop scheduler operates using a FIFO queue. After a job is partitioned into individual tasks, they are loaded into the queue and assigned to free slots as they become available on TaskTracker nodes. Although there is support for assignment of priorities to jobs, this is not turned on by default. Typically each job would use the whole cluster, so jobs had to wait for their turn. Even though a shared cluster offers great potential for offering large resources to many users, the problem of sharing resources fairly between users requires a better scheduler. Production jobs need to complete in a timely manner, while allowing users who are making smaller ad hoc queries to get results back in a reasonable time.

## 3.2 Fair Scheduler

The Fair Scheduler [7] was developed at Facebook to manage access to their Hadoop cluster and subsequently released to the Hadoop community. The Fair Scheduler aims to give every user a fair share of the cluster capacity over time. Users may assign jobs to pools, with each pool allocated a guaranteed minimum number of Map and Reduce slots. Free slots in idle pools may be allocated to other pools, while excess capacity within a pool is shared among jobs. The Fair Scheduler supports preemption, so if a pool has not received its fair share for a certain period of time, then the scheduler will kill tasks in pools running over capacity in order to give the slots to the pool running under capacity. In addition, administrators may enforce priority settings on certain pools. Tasks are therefore scheduled in an interleaved manner, based on their priority within their pool, and the cluster capacity and usage of their pool. As jobs have their tasks allocated to Task Tracker slots for computation, the scheduler tracks the deficit between the amount of time actually used and the ideal fair allocation for that job. As slots become available for scheduling, the next task from the job with the highest time deficit is assigned to the next free slot. Over time, this has the effect of ensuring each job receive roughly equal amounts of resources. Shorter jobs are allocated sufficient resources to finish quickly. At the same time, longer jobs are guaranteed to not be starved of resources.

## 3.3 Capacity Scheduler
Capacity Scheduler [3] originally developed at Yahoo addresses a usage scenario where the number of users is large, and there is a need to ensure a fair allocation of computation resources amongst users. The Capacity Scheduler allocates jobs based on the submitting user to queues with configurable numbers of Map and Reduce slots. Queues that contain jobs are given their configured capacity, while free capacity in a queue is shared among other queues. Within a queue, scheduling operates on a modified priority queue basis with specific user limits, with priorities adjusted based on the time a job was submitted, and the priority setting allocated to that user and class of job. When a Task Tracker becomes free, the queue with the lowest load is chosen, from which the oldest remaining job is chosen. A task is then scheduled from that job. Overall, this has the effect of enforcing cluster capacity sharing among users, rather than among jobs, as was the case in the Fair Scheduler.

## 4. SCHEDULER IMPROVEMENTS
Many researchers are working on opportunities for improving the scheduling policies in Hadoop. Recent efforts such as Delay Scheduler [9], Dynamic Proportional Scheduler [10] offer differentiated service for Hadoop jobs allowing users to adjust the priority levels assigned to their jobs. However, this does not guarantee that the job will be completed by a specific deadline. Deadline Constraint Scheduler [11] addresses the issue of deadlines but focuses more on increasing system utilization. The Schedulers described above attempt to allocate capacity fairly among users and jobs, they make no attempt to consider resource availability on a more fine-grained basis. Resource Aware Scheduler [12] considers the resource availability to schedule jobs. In the following sections we compare and contrast the work done by the researchers on various Schedulers.

## 4.1 Longest Approximate Time to End (LATE) - Speculative Execution
It is not uncommon for a particular task to continue to progress slowly. This may be due to several reasons like–high CPU load on the node, slow background processes etc. All tasks should be finished for completion of the entire job. The scheduler tries to detect a slow running task to launch another equivalent task as a backup which is termed as speculative execution of tasks. If the backup copy completes faster, the overall job performance is improved. Speculative execution is an optimization but not a feature to ensure reliability of jobs. If bugs cause a task to hang or slow down then speculative execution is not a solution, since the same bugs are likely to affect the speculative task also. Bugs should be fixed so that the task doesn't hang or slow down. The default implementation of speculative execution relies implicitly on certain assumptions: a) Uniform Task progress on nodes b) Uniform computation at all nodes. That is, default implementation of speculative execution works well on homogeneous clusters. These assumptions break down very easily in the heterogeneous clusters that are found in real-world production scenarios. Zaharia et al [13] proposed a modified version of speculative execution called Longest Approximate Time to End (LATE) algorithm that uses a different metric to schedule tasks for speculative execution. Instead of considering the progress made by a task so far, they compute the estimated time remaining, which gives a more clear assessment of a straggling tasks' impact on the overall job response time. They demonstrated significant improvements by Longest Approximate Time to End (LATE) algorithm over the default speculative execution.

## 4.2 Delay Scheduling
Fair scheduler is developed to allocate fair share of capacity to all the users. Two locality problems identified when fair sharing is followed are – head-of-line scheduling and sticky slots. The first locality problem occurs in small jobs (jobs that have small input files and hence have a small number of data blocks to read). The problem is that whenever a job reaches the head of the sorted list for scheduling, one of its tasks is launched on the next slot that becomes free irrespective of which node this slot is on. If the head-of-line job is small, it is unlikely to have data locally on the node that is given to it. Head-of-line scheduling problem was observed at Facebook in a version of HFS without delay scheduling. The other locality problem, sticky slots, is that there is a tendency for a job to be assigned the same slot repeatedly. The problems aroused because following a strict queuing order forces a job with no local data to be scheduled.

To overcome the Head of line problem, scheduler launches a task from a job on a node without local data to maintain fairness, but violates the main objective of MapReduce that schedule tasks near their input data. Running on a node that contains the data (node locality) is most efficient, but when this is not possible, running on a node on the same rack (rack locality) is faster than running off-rack. Delay scheduling is a solution that temporarily relaxes fairness to improve locality by asking jobs to wait for a scheduling opportunity on a node with local data. When a node requests a task, if the head-of-line job cannot launch a local task, it is skipped and looked at subsequent jobs. However, if a job has been skipped long enough, non-local tasks are allowed to launch to avoid starvation. The key insight behind delay scheduling is that although the first slot we consider giving to a job is unlikely to have data for it, tasks finish so quickly that some slot with data for it will free up in the next few seconds.

## 4.3 Dynamic Priority Scheduling

Thomas Sandholm *et al* [10] proposed Dynamic Priority Scheduler that supports capacity distribution dynamically among concurrent users based on priorities of the users. Automated capacity allocation and redistribution is supported in a regulated task slot resource market. This approach allows users to get Map or Reduce slot on a proportional share basis per time unit. These time slots can be configured and called as allocation interval. It is typically set to somewhere between 10 seconds and 1 minute. For example a max capacity of 15 Map slots gets allocated proportionally to three users. The central scheduler contains a Dynamic Priority Allocator and a Priority Enforcer component responsible for accounting and schedule enforcement respectively. This model appears to favor users with small jobs than users with bigger jobs. However Hadoop MapReduce supports scaling down of big jobs to small jobs to make sure that fewer concurrent tasks runs by consuming the same amount of resources.

To avoid starvation, queue blocking and to respond to user demand fluctuations more quickly preemption is also supported. In this mechanism task slots that were allocated may be preempted and allocated to other users if they were not used for long time. As a result of variable pricing mechanism users to get guaranteed slot during demand periods has to pay more. This scheme discourages the free-riding and gaming by users. However, the Hadoop MapReduce scheduling framework allows jobs to be split up in finer grained tasks that can run and possibly fail and recover independently. So the only thing the end users would need to worry about is to get a good enough average capacity over some time to meet their deadlines. This introduces the difficulty of making spending rate decisions to meet the SLA and deadline requirements. Possible starvation of low-priority (low-spending) tasks can be mitigated by using the standard approach in Hadoop of limiting the time each task is allowed to run on a node. Moreover, this new mechanism also allows administrators to set budgets for different users and let them individually decide whether the current price of preempting running tasks is within their budget or if they should wait until the current users run out of their budget. The fact that Hadoop uses task and slot level scheduling and allocation as opposed to job level scheduling also avoids many starvation scenarios. If there is no contention, i.e. there are enough slots available to run all tasks from all jobs submitted, the cost for excess resources essentially becomes free because of the work conserving principle of this scheduler. However, the guarantees of maintaining these excess resources are reduced. To see why, consider new users deciding whether to submit jobs or not. If they see that the price is high they may wait to preempt currently running jobs, but if the resources are essentially given out for free they are likely to lay claim on as many resources they can immediately. We note that the Dynamic Priority scheduler can easily be configured to mimic the behavior of the other schedulers. If no queues or users have any credits left the scheduler reduces to a FIFO scheduler. If all queues are configured with the same share (spending rate in our case) and the allocation interval is set to a very large value the scheduler reduces to the behavior of the static fair-share schedulers.

## 4.4 Deadline Constraint Scheduler

Deadline Constraint Scheduler [11] addresses the issue of deadlines but focuses more on increasing system utilization. Dealing with deadline requirements in Hadoop-based data processing is done by (1) a job execution cost model that considers various parameters like map and reduce runtimes, input data sizes, data distribution, etc., (2) a Constraint-Based Hadoop Scheduler that takes user deadlines as part of its input. Estimation model determines the available slot based a set of assumptions:

- All nodes are homogeneous nodes and unit cost of processing for each map or reduce node is equal
- Input data is distributed uniform manner such that each reduce node gets equal amount of reduce data to process
- Reduce tasks starts after all map tasks have completed;
- The input data is already available in HDFS.

Schedulability of a job is determined based on the proposed job execution cost model independent of the number of jobs running in the cluster. Jobs are only scheduled if specified deadlines can be met. After a job is submitted, schedulability test is performed to determine whether the job can be finished within the specified deadline or not. Free slots availability is computed at the given time or in the future irrespective of all the jobs running in the system. The job is enlisted for scheduling after it is determined that the job can be completed within the given deadline. A job is schedulable if the minimum number of tasks for both map and reduce is less than or equal to the available slots. This Scheduler shows that when a deadline for job is different, then the scheduler assigns different number of tasks to TaskTracker and makes sure that the specified deadline is met.

## 4.5 Resource Aware Scheduling

The Fair Scheduler [7] and Capacity Scheduler described above attempt to allocate capacity fairly among users and jobs without considering resource availability on a more fine-grained basis. As CPU and disk channel capacity has been increasing in recent years, a Hadoop cluster with heterogeneous nodes could exhibit significant diversity in processing power and disk access speed among nodes. Performance could be affected if multiple processor-intensive or data-intensive tasks are allocated onto nodes with slow processors or disk channels respectively. This possibility arises as the Job Tracker simply treats each Task Tracker node as having a number of available task "slots" Even the improved LATE speculative execution could end up increasing the degree of congestion within a busy cluster, if speculative copies are simply assigned to machines that are already close to maximum resource utilization.

Resource Aware Scheduling in Hadoop has become one of the Research Challenges [14][15] in Cloud Computing. Scheduling in Hadoop is centralized, and worker initiated. Scheduling decisions are taken by a master node, called the JobTracker, whereas the worker nodes, called TaskTrackers are responsible for task execution. The JobTracker maintains a queue of currently running jobs, states of TaskTrackers in a cluster, and list of tasks allocated to each TaskTracker. Each Task Tracker node is currently configured with a maximum number of available computation slots. Although this can be configured on a per-node basis to reflect the actual processing power and disk channel speed, etc available on cluster machines, there is no online modification of this slot capacity available. That is, there is no way to reduce congestion on a machine by advertising a reduced capacity. In this mechanism, each Task Tracker node monitors resources such as CPU utilization, disk channel IO in bytes/s, and the number of page faults per unit time for the memory subsystem. Although we anticipate that other metrics will prove useful, we propose these as the basic three resources that must be tracked at all times to improve the load balancing on cluster machines. In particular, disk channel loading can significantly impact the data loading and writing portion of Map and Reduce tasks, more so than the amount of free space available. Likewise, the inherent opacity of a machine's virtual memory management state means that monitoring page faults and virtual memory-induced disk thrashing is a more useful indicator of machine load than simply tracking free memory.

Two possible resource-aware Job Tracker scheduling mechanisms are: 1) *Dynamic Free Slot Advertisement*-Instead of having a fixed number of available computation slots configured on each Task Tracker node, this number is computed dynamically using the resource metrics obtained from each node. In one possible heuristic, overall resource availability is set on a machine to be the minimum availability across all resource metrics. In a cluster that is not running at maximum utilization at all times, this is expected to improve job response times significantly as no machine is running tasks in a manner that runs into a resource bottleneck. 2) *Free Slot Priorities/Filtering*- In this mechanism, cluster administrators will configure maximum number of compute slots per node at configuration time. The order in which free TaskTracker slots are advertised is decided according to their resource availability. As TaskTracker slots become free, they are buffered for some small time period (say, 2s) and advertised in a block. TaskTracker slots with higher resource availability are presented first for scheduling tasks on. In an environment where even short jobs take a relatively long time to complete, this will present significant performance gains. Instead of scheduling a task onto the next available free slot (which happens to be a relatively resource-deficient machine at this point), job response time would be improved by scheduling it onto a resource-rich machine, even if such a node takes a longer time to become available. Buffering the advertisement of free slots allowed for this scheduling allocation.

## 5. CONCLUSION & FUTURE WORK
Ability to make Hadoop scheduler resource aware is one the emerging research problem that grabs the attention of most of the researchers as the current implementation is based on statically configured slots. This paper summarizes pros and cons of Scheduling policies of various Hadoop Schedulers developed by different communities. Each of the Scheduler considers the resources like CPU, Memory, Job deadlines and IO etc. All the schedulers discussed in this paper addresses one or more problem(s) in scheduling in Hadoop. Nevertheless all the schedulers discussed above assumes homogeneous Hadoop clusters. Future work will consider scheduling in Hadoop in Heterogeneous Clusters.